\documentclass[journal]{IEEEtran}

\usepackage{amsmath}
\usepackage{amsthm}
\usepackage{amsfonts}
\usepackage{amssymb}
\usepackage{epsfig}
\usepackage{psfrag}
\usepackage{graphicx}
\usepackage{float}
\usepackage{multirow}
\usepackage{tabularx}
\usepackage{color}
\usepackage{algorithm}
\usepackage{algorithmic}
\usepackage{lineno}
\usepackage{hyperref}
\usepackage{multicol}
\usepackage{commath}
\usepackage{stfloats}
\usepackage{wrapfig}
\usepackage{subcaption}
\usepackage{caption}
\usepackage{cleveref}
\usepackage{tikz}
\usetikzlibrary{shapes.geometric, arrows}
\usepackage{adjustbox}
\usepackage{pifont}
\usepackage[sort&compress,square,sort,comma,numbers]{natbib}
\usepackage{soul}

\newcolumntype{C}[1]{>{\centering\let\newline\\\arraybackslash\hspace{-2mm}}m{#1}}

\tikzstyle{startstop} = [rectangle, rounded corners, minimum width=2cm, minimum height=1cm, text centered, draw=black, fill=white]
\tikzstyle{io} = [trapezium, trapezium left angle=70, trapezium right angle=110, minimum width=2cm, minimum height=1cm, text width=4cm, text centered, draw=black, fill=white]
\tikzstyle{process} = [rectangle, minimum width=2cm, minimum height=1cm, text width=3cm, text centered, draw=black, fill=white]
\tikzstyle{decision} = [diamond, minimum width=2cm, minimum height=1cm, text width=2cm, text centered, draw=black, fill=white]
\tikzstyle{arrow} = [thick,->,>=stealth]

\begin{document}

\date{}

\title{Reinforcing Localization Credibility Through Convex Optimization}

\author{Slavisa~Tomic,
        Marko~Beko,
        Yakubu Tsado,
        Bamidele~Adebisi
        and Abiola~Oladipo
\thanks{This research was partially funded by the European Union’s Horizon Europe Research and Innovation Programme under the Marie Sk\l{}odowska-Curie grant agreement No. 101086387 and by Funda\c{c}\~{a}o para a Ci\^{e}ncia e a Tecnologia under Projects UIDB/04111/2020, UIDB/50008/2020, ROBUST EXPL/EEI-EEE/0776/2021, 2021.04180.CEECIND, and by the UK Engineering and Physical Sciences Research Council (EPSRC) and Horizon Europe Guarantee [Grant number EP/X039021/1]- REMARKABLE.}
\thanks{S.T. is with Lusofona University, Campo Grande 376, 1749 - 024 Lisboa, Portugal (e-mail: slavisa.tomic@ulusofona.pt). M.B. is with Instituto de Telecomunica\c{c}\~{o}es, Instituto Superior T\'{e}cnico, Universidade de Lisboa, 1049-001 Lisbon, Portugal (e-mail: marko.beko@tecnico.ulisboa.pt). S.T. and M.B. are also with and Center of Technology and Systems (UNINOVA-CTS) and Associated Lab of Intelligent Systems (LASI), 2829-516 Caparica, Portugal.
Y.T. and B.O. are with Department of Engineering, Manchester Metropolitan University, Manchester, UK (e-mail: Y.Tsado@mmu.ac.uk, B.Adebisi@mmu.ac.uk). A.O. is with Tiwakiki Consulting Limited, 2, Mandley Avenue, Manchester, UK (e-mail: abiola.oladipo@tiwakikiconsulting.com).}}

\maketitle



\begin{abstract}
This work proposes a novel approach to reinforce localization security in wireless networks in the presence of malicious nodes that are able to manipulate (spoof) radio measurements. It substitutes the original measurement model by another one containing an auxiliary variance dilation parameter that disguises corrupted radio links into ones with large noise variances. This allows for relaxing the non-convex maximum likelihood estimator (MLE) into a semidefinite programming (SDP) problem by applying convex-concave programming (CCP) procedure. The proposed SDP solution simultaneously outputs target location and attacker detection estimates, eliminating the need for further application of sophisticated detectors. Numerical results corroborate excellent performance of the proposed method in terms of localization accuracy and show that its detection rates are highly competitive with the state of the art.
\end{abstract}




\section{Introduction}
\label{sec:intro}

Reliably determining locations of devices in wireless networks has always been a challenging task and will play an essential role in the forthcoming 6G wireless systems, where manipulated (spoofed) location information can lead to catastrophic repercussions. Even though secure non-satellite based localization problem has been attracting interest in recent years~\cite{Li:2005, Liu:2007, Garg:2012, Liu:2019, Li:2021, Beko:2021, Tomic_TVT:2022, Mukhopadhyay:2018, Mukhopadhyay:2021, Tomic_TVT1:2024, Tomic_TVT2:2024}, there is no uniquely accepted solution and there is still potential for improvement in all relevant aspects (localization accuracy, detection rates and complexity).

Several studies have explored secure localization techniques to mitigate the impact of malicious anchors. Early approaches include least median of squares (LMS) and radio frequency (RF) fingerprinting, as investigated in~\cite{Li:2005}. The LMS method selects the best localization estimate based on median residues, while RF fingerprinting employs a median-based distance metric. Attack-resistant methods such as attack-resistant minimum mean square estimation (ARMMSE) and voting scheme (VS) were proposed in~\cite{Liu:2007} to detect and exclude malicious anchors by analyzing inconsistencies or leveraging grid-based voting mechanisms. An iterative gradient descent approach for handling spoofing attacks was introduced in~\cite{Garg:2012}, where residues from malicious anchors tend to be higher and can thus be filtered out to enhance location estimation. More recent approaches employ weighted least squares (WLS) estimator~\cite{Mukhopadhyay:2018} and density-based spatial clustering to distinguish normal from contaminated location points, as seen in~\cite{Liu:2019}, combined with sequential probability ratio tests for anchor authentication. Probabilistic models such as maximum a posteriori (MAP) estimators, solved using variational message passing, have been applied for secure localization in mobile wireless sensor networks in~\cite{Li:2021}. Other methods, such as the weighted central mass (WCM) approach in~\cite{Beko:2021}, exploit geometric properties to obtain initial estimates and iteratively refine them using trust region sub-problems (GTRS). A generalization of this method was presented in~\cite{Tomic_TVT:2022}, where a generalized likelihood ratio test (GLRT) was used for detection, and the law of cosines (LC) was applied to reformulate the problem into a GTRS. Recent advancements include robust optimization techniques such as secure weighted least squares (SWLS) and $l_1$-norm-based approaches, introduced in~\cite{Mukhopadhyay:2021}, which use statistical deviation thresholds and clustering methods to detect malicious nodes. Additionally, robust formulations using min-max optimization, second-order cone programming (SOCP), and robust GTRS (R-GTRS) were proposed in~\cite{Tomic_TVT1:2024} to enhance localization security by treating attacks as nuisance parameters. Another approach, based on the alternating direction method of multipliers (ADMM) that employs least squares criterion together with a decomposition-coordination iterative scheme to refine localization estimates was presented in~\cite{Tomic_TVT2:2024}. A VS based on the distance to hyperplanes formed by pairs of anchors was introduced in~\cite{Tomic_IoT:2025}, with attacker classification being performed by confidence intervals.

In huge contrast to the existing solutions, this work takes a radically different approach and proposes a substitution of the original measurement model by a different one that accounts for noise variance dilation in order to capture corrupted measurements as ones with large noise variances. This is achieved by defining an auxiliary variance amplification parameter which allows for transformation of the non-convex maximum likelihood estimator (MLE) into a convex semidefinite programming (SDP) problem by applying a simple convex-concave programming (CCP) procedure. The proposed SDP solution not only sets a new achievable lower bound on localization accuracy, but also holds within attacker detection estimation, depriving the need for refined detectors.



\section{Problem Formulation}
\label{sec:problem_formulation}

Consider a $q$-dimensional ($q = 2$ or $q = 3$) wireless network, composed of two types of nodes: targets (nodes whose locations one desires to determine) and anchors (nodes whose locations are known and are exploited as reference points for localization purposes). Targets communicate with anchors that estimate distance information from the received signal (e.g., via time of arrival or received signal strength), which is then exploited for localization process of a single target at a time. The true location of a target is denoted by $\boldsymbol{x}$, while the true locations of anchors are denoted by $\boldsymbol{a}_i, i=1,...,N$. Finally, a fraction of the anchors is considered corrupted and attempts to hinder the localization process by modifying its distance measurements, i.e., by performing spoofing attacks. 


One can formulate secure range-based localization problem as a system of non-linear equations, where each equation corresponds to measured distance between an anchor and a target, corrupted with noise and potentially by a spoofing attack~\cite{Tomic_TVT:2022}. Hence, the $k$-th distance measurement sample, with $1 \leq k \leq K$, between the target and the $i$-th anchor (in meters) can be modelled~\cite{Tippenhauer:2011}-\cite{Seo:2021} as
\begin{equation}
d_{i,k} = \|\boldsymbol{x}-\boldsymbol{a}_i\| + \delta_i + n_{i,k},
\label{eq:model}
\end{equation}
where $n_{i,k}$ stands for the measurement noise, modelled as a zero-mean Gaussian random variable with variance $\sigma_{i,k}^2$, i.e., $n_{i,k} \sim \mathcal{N}(0, \sigma_{i,k}^2)$, and $\delta_i \in \mathbb{R}$ represents the (unknown) intensity of the spoofing attack. It is clear that when $\delta_i = 0$ the $i$-th anchor is genuine (i.e., $i \in \mathcal{G}$), while $\delta_i \neq 0$ corresponds to the $i$-th anchor being corrupted (i.e., $i \in \mathcal{C}$).

%

For simplicity and with no loss of generality in the following derivations, the median of the $K$ range measurements, $d_i$, is employed as the observation at the $i$-th anchor; thus, the subscript $k$ is omitted. Likewise, the noise variances are assumed equal for all links (and samples), i.e., $\sigma_1^2 = \hdots = \sigma_N^2 = \sigma^2$.

\begin{figure}
\centering
\hspace*{-0mm}\includegraphics[width=.75\linewidth]{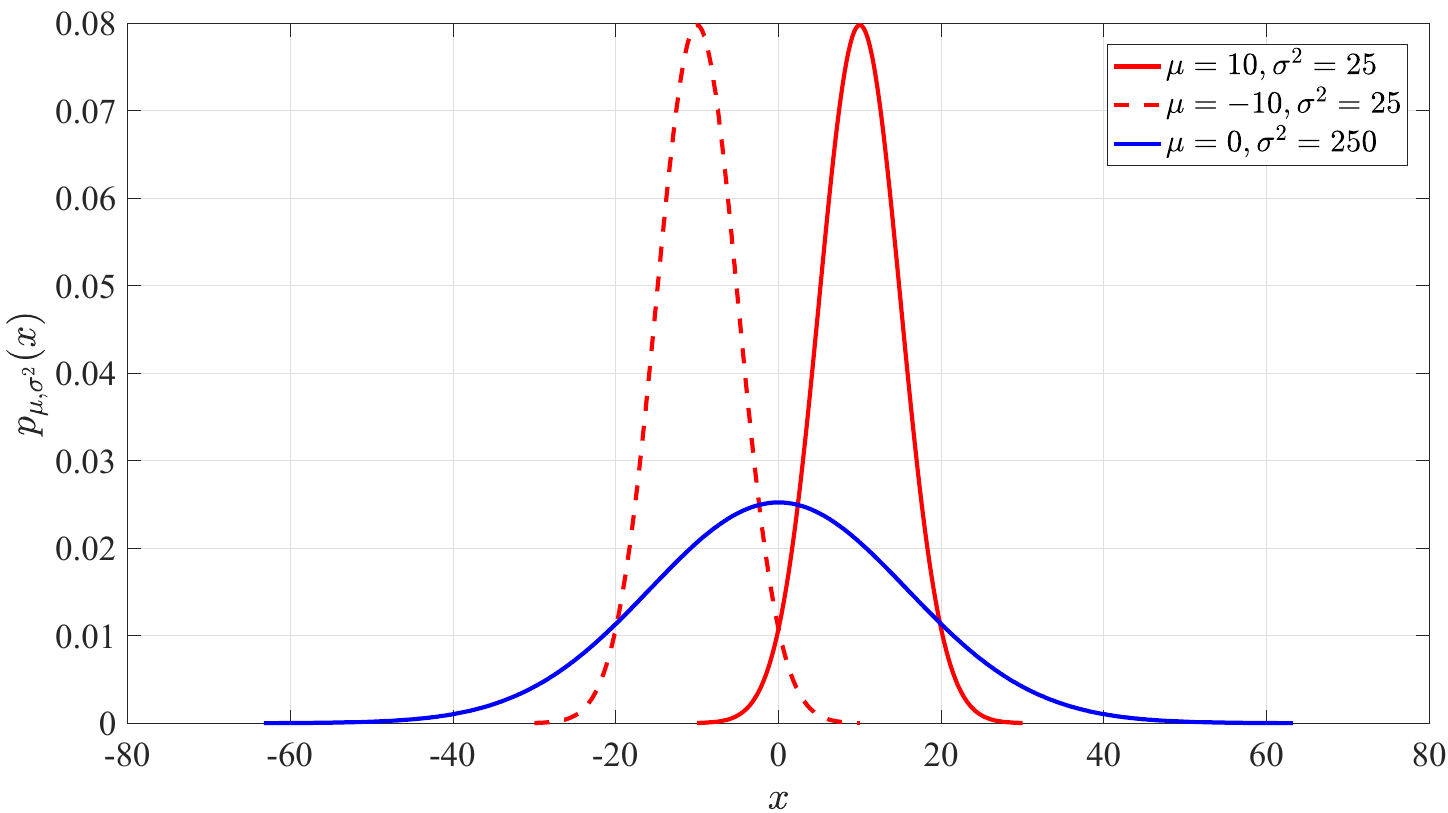}
\vspace*{-2mm}\caption{Dilating noise variance to circumvent spoofing.}
\label{fig:switching_pdfs}
\end{figure}

From~\eqref{eq:model} and following maximum likelihood principle~\cite{Kay:1993}, one can formulate the localization problem as an optimization problem with $\boldsymbol{x}$ and $\delta_i$ as variables~\cite{Tomic_TVT2:2024}. Nevertheless, such a problem is non-convex~\cite{Soares:2020}, since the argument of the square factor $(\|\boldsymbol{x}-\boldsymbol{a}_i\| +\delta_i - d_{i})^2$ has a negative region when $\|\boldsymbol{x}-\boldsymbol{a}_i\| + \delta_i < d_{i}$, and is under-determined ($N+q$ unknowns with $N$ equations). Hence, this work circumvents the presence of spoofed measurements by dilating noise variance, as illustrated in Fig.~\ref{fig:switching_pdfs}. To do so, the model in~\eqref{eq:model} is rewritten as
\begin{equation}
d_{i} = \|\boldsymbol{x}-\boldsymbol{a}_i\| + \epsilon_{i},
\label{eq:new_model}
\end{equation}
with $\epsilon_i \sim \mathcal{N}(0, \rho_i \sigma^2)$ and $\rho_i = \begin{cases}
1, & \text{if } i \in \mathcal{G}\\
R_i > 1, & i \in \mathcal{C}
\end{cases}$.

Note that the model in~\eqref{eq:new_model} is adopted to allow for the following derivations; the actual observations still come from~\eqref{eq:model}.

In what follows, this approximation is justified by showing that the Cramér-Rao lower bound (CRLB) of an estimator resulting from~\eqref{eq:new_model} is approximately equal to the one from~\eqref{eq:model}. Let $\boldsymbol{\theta} = [\boldsymbol{x}^T, \nu_1, \hdots, \nu_N]^T$ denote the $N+q$ vector of all unknown parameters, i.e., $\nu_i = \delta_i$ for~\eqref{eq:model} and $\nu_i = \rho_i$ for~\eqref{eq:new_model}, $i = 1, \hdots, N$. According to~\cite{Kay:1993}, the variance of any unbiased estimator is lower bounded by $\text{var}(\hat{\theta}_i) \geq [\boldsymbol{F}^{-1}(\boldsymbol{\theta})]_{ii}$, with $\boldsymbol{F}(\boldsymbol{\theta})$ being the $(N+q)\times (N+q)$ Fisher information matrix, whose elements are defined as $[F(\boldsymbol{\theta})]_{ij} = -\mathbb{E}\left[ \frac{\partial^2 L(\boldsymbol{d}; \boldsymbol{\theta})}{\partial \theta_i \partial\theta_j } \right]$, where $i, j = 1, \hdots, N+q$ and $L(\boldsymbol{d}; \boldsymbol{\theta})$ is the log-likelihood of $\boldsymbol{d} = [d_1, \hdots, d_N]$ parametrized by $\boldsymbol{\theta}$. One can partition $\boldsymbol{F}(\boldsymbol{\theta})$ as
\begin{equation}
\boldsymbol{F}(\boldsymbol{\theta}) =
\begin{bmatrix}
\boldsymbol{A}(\boldsymbol{\theta}) \in \mathbb{R}^{q \times q} & \boldsymbol{B}(\boldsymbol{\theta}) \in \mathbb{R}^{q \times N}\\
\boldsymbol{B}^T(\boldsymbol{\theta}) \in \mathbb{R}^{N \times q} & \boldsymbol{C}(\boldsymbol{\theta}) \in \mathbb{R}^{N \times N}
\end{bmatrix},
\nonumber
\end{equation}
such that $\boldsymbol{A}(\boldsymbol{\theta}) = \frac{1}{\sigma^2} \sum_{i=1}^N \frac{(\boldsymbol{x} - \boldsymbol{a}_i)(\boldsymbol{x} - \boldsymbol{a}_i)^T}{\| \boldsymbol{x} - \boldsymbol{a}_i \|}$, $\boldsymbol{B}(\boldsymbol{\theta}) = \frac{1}{\sigma^2}[\frac{(\boldsymbol{x} - \boldsymbol{a}_1)}{\| \boldsymbol{x} - \boldsymbol{a}_1 \|}, \hdots, \frac{(\boldsymbol{x} - \boldsymbol{a}_N)}{\| \boldsymbol{x} - \boldsymbol{a}_N \|}]$, $\boldsymbol{C}(\boldsymbol{\theta}) = \frac{1}{\sigma^2} \boldsymbol{I}_{N}$ for~\eqref{eq:model}, and $\boldsymbol{A}(\boldsymbol{\theta}) = \frac{1}{\sigma^2} \sum_{i=1}^N \frac{(\boldsymbol{x} - \boldsymbol{a}_i)(\boldsymbol{x} - \boldsymbol{a}_i)^T}{\rho_i\| \boldsymbol{x} - \boldsymbol{a}_i \|}$, $\boldsymbol{B}(\boldsymbol{\theta}) = \boldsymbol{0}_{2 \times N}$, $\boldsymbol{C}(\boldsymbol{\theta}) = \text{diag}([-\frac{1}{\rho_1^2}, \hdots, -\frac{1}{\rho_N^2}])$ for~\eqref{eq:new_model}. The CRLB for $\boldsymbol{x}$ can then be derived as $\text{var}(\boldsymbol{\hat{x}}) \geq \text{trace}(\boldsymbol{F}_{\boldsymbol{x}}(\boldsymbol{\theta}))$, by computing the Schur complement of $\boldsymbol{C}(\boldsymbol{\theta})$, i.e.,
\begin{equation}
\boldsymbol{F}_{\boldsymbol{x}}(\boldsymbol{\theta}) = \boldsymbol{A}(\boldsymbol{\theta}) - \boldsymbol{B}(\boldsymbol{\theta}) \boldsymbol{C}^{-1}(\boldsymbol{\theta}) \boldsymbol{B}^T(\boldsymbol{\theta}).
\nonumber
\end{equation}
Recall that $\delta_i = 0$ and $\rho_i = 1$ when $i \in \mathcal{G}$; thus, it is clear that $\boldsymbol{F}_{\boldsymbol{x}} = \frac{1}{\sigma^2} \sum_{i=1}^N \frac{(\boldsymbol{x} - \boldsymbol{a}_i)(\boldsymbol{x} - \boldsymbol{a}_i)^T}{\| \boldsymbol{x} - \boldsymbol{a}_i \|}$ for the two models, while when $i \in \mathcal{C}$, the CRLBs are approximately equal for $\rho \gg 1$.

According to~\eqref{eq:new_model}, one can cast the localization problem as
\begin{equation}
(\widehat{\boldsymbol{x}}, \widehat{\boldsymbol{\rho}}) = \underset{\boldsymbol{x}, \boldsymbol{\rho}}{\text{\text{arg\,max}}} \,\,\, L(\boldsymbol{x}, \boldsymbol{\rho})
\label{eq:likelihood}
\end{equation}
where $\boldsymbol{\rho} = [\rho_1, \hdots, \rho_N]^T$ and $L(\boldsymbol{x}, \boldsymbol{\rho}) = -\displaystyle\sum_{i=1}^N \ln\left\{ \frac{1}{\sqrt{2\pi\rho_i\sigma^2}} \exp\left\{ -\frac{(d_{i} - \|\boldsymbol{x}-\boldsymbol{a}_i\|)^2}{2\rho_i\sigma^2} \right\} \right\}$ is the log-likelihood function. Then, since
\begin{equation}
L(\boldsymbol{x}, \boldsymbol{\rho}) \propto \displaystyle\sum_{i=1}^N \frac{(d_{i} - \|\boldsymbol{x}-\boldsymbol{a}_i\|)^2}{\rho_i} + \displaystyle\sum_{i=1}^N \ln(\rho_i)
\nonumber
\end{equation}
and by relaxing the binary variable $\rho_i$ into a continuous variable in the range $[1, \infty)$, yields
\begin{subequations}
\begin{equation}
\underset{\boldsymbol{x}, \boldsymbol{\rho}}{\text{minimize}} \, \displaystyle\sum_{i=1}^N \frac{(d_{i} - \|\boldsymbol{x}-\boldsymbol{a}_i\|)^2}{\rho_i} + \displaystyle\sum_{i=1}^N \ln(\rho_i)
\label{eq:loc_probl_cost_fun}
\end{equation}
\begin{equation}
\text{subject\,\,to} \,\,\, \rho_i \geq 1, i = 1, \hdots, N.
\label{eq:loc_probl_constraint}
\end{equation}
\label{eq:loc_problem}
\end{subequations}

The problem in~\eqref{eq:loc_problem} involves an objective function that is a sum of non-convex and concave terms and is thus difficult to tackle directly. Nevertheless, the following section shows how this problem can be converted into a convex, SDP, problem.



\section{The Proposed SDP Estimator}
\label{sec:proposed_method}

Current section offers details on the derivation of the proposed algorithm for credible localization. Start by defining auxiliary variables $y_i = d_{i} - \|\boldsymbol{x}-\boldsymbol{a}_i\|$, $\boldsymbol{Y} = \boldsymbol{y} \boldsymbol{y}^T$ and $\boldsymbol{X} = \boldsymbol{x} \boldsymbol{x}^T$, where $\boldsymbol{y} = [y_1, \hdots, y_N]^T$. Then, it follows that
\begin{equation}
(y_i - d_{i})^2\hspace*{-1mm}=\hspace*{-1mm}\|\boldsymbol{x}-\boldsymbol{a}_i\|^2\hspace*{-1mm}\Leftrightarrow\hspace*{-1mm}Y_{ii} - 2d_iy_i+d_i^2\hspace*{-1mm}=\hspace*{-1mm}\text{tr}(\boldsymbol{X}) - 2\boldsymbol{a}_i^T \boldsymbol{x} + \|\boldsymbol{a}_i\|^2
\nonumber
\end{equation}
and consequently
\begin{equation}
(y_i - d_{i}) (y_j - d_{j}) = \|\boldsymbol{x}-\boldsymbol{a}_i\| \|\boldsymbol{x}-\boldsymbol{a}_j\|,
\nonumber
\end{equation}
which, by applying the Cauchy–Schwarz inequality, yields
\begin{equation}
Y_{ij}\hspace*{-0.5mm}\geq \abs{\text{tr}(\boldsymbol{X})\hspace*{-0.5mm}-\hspace*{-0.5mm}\boldsymbol{a}_j^T \boldsymbol{x} - \boldsymbol{a}_i^T \boldsymbol{x} + \boldsymbol{a}_j^T \boldsymbol{a}_i}\hspace*{-0.5mm}+\hspace*{-0.5mm}d_jy_i\hspace*{-0.5mm}+\hspace*{-0.5mm}d_iy_j\hspace*{-0.5mm}-\hspace*{-0.5mm}d_id_j.
\nonumber
\end{equation}

Moreover, introduce $e_i = \frac{(d_{i} - \|\boldsymbol{x}-\boldsymbol{a}_i\|)^2}{\rho_i} = \frac{(y_i)^2}{\rho_i} = \frac{Y_{ii}}{\rho_i}$ as an epigraph variable. This substitution, together with $\rho_i \geq 1$ results in
\begin{equation}
e_i \geq 0, \,\, \text{and} \,\, Y_{ii} \geq e_i.
\nonumber
\end{equation}
It also allows to deal with the second (concave term) summation of~\eqref{eq:loc_probl_cost_fun} efficiently through an iterative CCP~\cite{Shen:2016} as follows. Note first that $\ln(\rho_i) = \underbrace{\ln(Y_{ii})}_{\text{concave}} \underbrace{-\ln(e_i)}_{\text{convex}}$. Thus, for a feasible initial point, $\hat{Y}_{ii}^{(0)}$, the concave terms are \emph{convexified} in iteration $t$ by an affine approximation (gradient evaluation) around that point, i.e.,
\begin{equation}
\ln\left(Y_{ii}; \hat{Y}_{ii}^{(t)}\right) = \ln\left(\hat{Y}_{ii}^{(t)}\right) + \frac{\left( Y_{ii} - \hat{Y}_{ii}^{(t)} \right)}{\hat{Y}_{ii}^{(t)}},
\nonumber
\end{equation}
for $t = 0, \hdots, T$, with $T$ being the maximum number of steps.

Therefore, by joining all described steps together, one arrives at the proposed SDP as
\begin{subequations}
\begin{equation}
\underset{\boldsymbol{x}, \boldsymbol{y}, \boldsymbol{X}, \boldsymbol{Y}, \boldsymbol{e}}{\text{minimize}} \,\, \displaystyle\sum_{i=1}^N e_i - \displaystyle\sum_{i=1}^N \ln(e_i) + \displaystyle\sum_{i=1}^N \frac{Y_{ii}}{\hat{Y}_{ii}^{(t)}}
\label{eq:SDP_cost_fun}
\end{equation}
\begin{equation}
\begin{array}{lcc}
\text{subject\,\,to} & e_i \geq 0, \,\, i = 1, \hdots, N,
\end{array}
\label{eq:SDP_const1}
\end{equation}
\begin{equation}
Y_{ii} \geq e_i, \,\, i = 1, \hdots, N,
\label{eq:SDP_const2}
\end{equation}
\begin{equation}
Y_{ii} = \text{tr}(\boldsymbol{X}) - 2\boldsymbol{a}_i^T \boldsymbol{x} + \|\boldsymbol{a}_i\|^2 + 2d_iy_i - d_i^2, \,\, i = 1, \hdots, N,
\label{eq:SDP_const3}
\end{equation}
\begin{equation}
\begin{array}{r}
Y_{ij}\hspace*{-0.5mm}\geq \abs{\text{tr}(\boldsymbol{X})\hspace*{-0.5mm}-\hspace*{-0.5mm}\boldsymbol{a}_j^T \boldsymbol{x} - \boldsymbol{a}_i^T \boldsymbol{x} + \boldsymbol{a}_j^T \boldsymbol{a}_i}\hspace*{-0.5mm}+\hspace*{-0.5mm}d_jy_i\hspace*{-0.5mm}+\hspace*{-0.5mm}d_iy_j\hspace*{-0.5mm}-\hspace*{-0.5mm}d_id_j,\\
i = 1, \hdots, N, \,\, j = 1, \hdots, N,
\end{array}
\label{eq:SDP_const4}
\end{equation}
\begin{equation}
\begin{bmatrix}
\boldsymbol{Y} & \boldsymbol{y}\\
\boldsymbol{y}^T & 1
\end{bmatrix} \succeq \boldsymbol{0}_{N+1}, \,\,
\begin{bmatrix}
\boldsymbol{X} & \boldsymbol{x}\\
\boldsymbol{x}^T & 1
\end{bmatrix} \succeq \boldsymbol{0}_{q+1},
\label{eq:SDP_const5}
\end{equation}
\label{eq:SDP}%
\end{subequations}
where all terms having no influence on minimization are omitted, the constraints $\boldsymbol{Y} = \boldsymbol{y} \boldsymbol{y}^T$ and $\boldsymbol{X} = \boldsymbol{x} \boldsymbol{x}^T$ are relaxed and written as semidefinite cone (SDC) constraints by applying the Schur complement, whereas the respective rank-1 constraints are dropped. The problem in~\eqref{eq:SDP} is an SDP and can be readily solved by CVX~\cite{CVX}.

The proposed solution for secure localization is summarized as a pseudo-code in Algorithm~\ref{al:SDP}. It is worth mentioning that the initial $\hat{\boldsymbol{Y}}^{(0)}$ is chosen assuming that all measurements are genuine. Given this, and since $Y_{ii} = (d_{i} - \|\boldsymbol{x}-\boldsymbol{a}_i\|)^2$ by construction, $\hat{Y}_{ii}^{(0)}$ is initially set to a relatively low value.
\begin{algorithm}\footnotesize
\caption{~Pseudo-code for the Proposed SDP Algorithm}
\begin{algorithmic}[1]
\REQUIRE $N:$ Number of anchors in the network
\REQUIRE $\boldsymbol{a}_i:$ True anchor locations $i = 1, ..., N$
\REQUIRE $K:$ Number of measurement samples
\REQUIRE $d_{i,k}:$ $k$-th distance measurement sample at $i$-th anchor
\REQUIRE $T:$ Maximum number of iterations
\REQUIRE $\tau:$ Threshold for stopping criterion
\STATE \textbf{Set:} $\hat{\boldsymbol{Y}}^{(0)} = 0.1 \times \boldsymbol{I}_N$, $\hat{\boldsymbol{x}}^{(0)} = 10^6 \times \boldsymbol{1}_{q \times 1}$, $\hat{\boldsymbol{x}}^{(1)} = \boldsymbol{0}_{q \times 1}$, $\hat{\mathcal{M}} = \varnothing$ and $t = 1$
\WHILE{$t \leq T$ and $\| \hat{\boldsymbol{x}}^{(t)} - \hat{\boldsymbol{x}}^{(t-1)} \| > \tau$}
	\STATE \textbf{Solve:}~\eqref{eq:SDP}
	\STATE \textbf{Update:}~$\hat{\boldsymbol{Y}}^{(t)} := \hat{\boldsymbol{Y}}$
	\STATE \textbf{Update:}~$\hat{\boldsymbol{x}}^{(t)} := \hat{\boldsymbol{x}}$
	\STATE \textbf{Update:}~$t := t + 1$
\ENDWHILE
\FOR{$i = 1, \hdots, N$}
	\IF{$\frac{\hat{y}_i^2}{e_i} > 1$}
		\STATE \textbf{Detect attacker:} $\hat{\mathcal{M}} := \hat{\mathcal{M}} \cup \left\{ i \right\}$
	\ENDIF
\ENDFOR
\STATE \textbf{Return:} $\hat{\boldsymbol{x}}$ and $\hat{\mathcal{M}}$
\end{algorithmic}
\label{al:SDP}
\end{algorithm}



\section{Performance Analysis}
\label{sec:performance}

In this section, the performance of the proposed solution is validated from various perspectives, including computational complexity, localization accuracy and detection rates.

\subsection{Complexity Analysis}
\label{complexity}

The worst case computational complexity of an SDP is calculated~\cite{Polik:2010} as
\begin{equation}
\mathcal{O}\left( C \left( m \displaystyle\sum_{i=1}^{N_{sd}} \left(n_i^{sd}\right)^3 + m^2 \displaystyle\sum_{i=1}^{N_{sd}} \left(n_i^{sd}\right)^2 + m^3 \right) \right),
\nonumber
\end{equation}
where $C$ is the iteration complexity, $m$ denotes the number of equality constraint, $n_i^{sd}$ is the dimension of the $i$-th SDC and $N_{sd}$ is the number of SDC constraints.

Assuming that $B_{\text{max}}$ and $B_{\text{ADMM}}$ stand for the maximum number of iterations for the GTRS-based and for the ADMM-based algorithms respectively, the worst-case computational complexity of the considered algorithms is summarized in Table~\ref{table:table1}. The table shows that the proposed solution is computationally the most burdensome, but this is compensated in accuracy, as will be seen in the following subsection.

\begin{table}\footnotesize
\caption{Summary of the Considered Algorithms}
\vspace*{-2mm}
  \begin{center}
	\small
	\begin{tabular}{|c|c|}
	\hline
	\textbf{Algorithm} & \textbf{Complexity}\\ \hline \hline
	SDP in Section~\ref{sec:proposed_method} & $\mathcal{O}(T \times N^{4.5})$\\ \hline
	LC-GTRS in~\cite{Tomic_TVT:2022}  & $2 \times \mathcal{O}(B_{\text{max}} \times N)$\\ \hline
    WLS in~\cite{Mukhopadhyay:2018}  & $\mathcal{O}(N)$\\ \hline
    LN-1 in~\cite{Mukhopadhyay:2021}  & $\mathcal{O}(B_{\text{ADMM}} \times N)$\\ \hline
    R-GTRS in~\cite{Tomic_TVT1:2024}  & $\mathcal{O}(B_{\text{{max}}} \times N)$\\ \hline
    WADMM in~\cite{Tomic_TVT2:2024}  & $\mathcal{O}(B_{\text{ADMM}} \times N)$\\ \hline
	\end{tabular}
   \end{center}
\label{table:table1}
\end{table}

\subsection{Localization and Detection Assessment}
\label{results}

The following figures disclose numerical results where $N$ anchors and a single target (at a time) were randomly deployed within a two-dimensional area of $B\times B~\text{m}^2$, with $B = 100$. Moreover, at most $N$/2 anchors were randomly chosen as corrupted, and this selection process was repeated $N_C=20$ times for each node deployment, considering $N_D=1000$ node deployments in total. All measurements were generated according to~\eqref{eq:model}, where $K=10$ measurement samples were considered and attacks were modelled through an exponential distribution whose rate was drawn from a uniform distribution on the interval $[0, \Delta]$ (m), i.e., $\delta_i \sim \pm \mathcal{E}(\mathcal{U}[0,\Delta])$, $\forall i$, where $\pm$ represents a random sign attribution to $\delta_i$. Results of the proposed SDP were achieved in maximum $T=3$ iterations with the threshold set at $\tau = B/200$ m. Note that the results could be further improved by fine-tuning $T$ and $\tau$, but our experiments have shown that its performance for the adopted values is sufficient. Lastly, all methods were tested against exactly the same noise and attack realizations to guarantee a fair comparison.

The main performance metric for localization is the root mean square error (RMSE), defined as $\text{RMSE} = \sqrt{\sum_{m=1}^{M_c} \frac{\|\boldsymbol{x}_{m} - \widehat{\boldsymbol{x}}_{m}\|^2}{M_C}},$ where $\widehat{\boldsymbol{x}}_{m}$ is the estimate of the true target location, $\boldsymbol{x}_{m}$, in the $m$-th Monte Carlo, $M_C = N_{D} \times N_{C}$, run, while detection is validated through the probability of correct detection, $P_{CD}$.

\begin{figure*}
\begin{center}
\begin{subfigure}{.325\textwidth}
\hspace*{-0mm}\includegraphics[width=\textwidth]{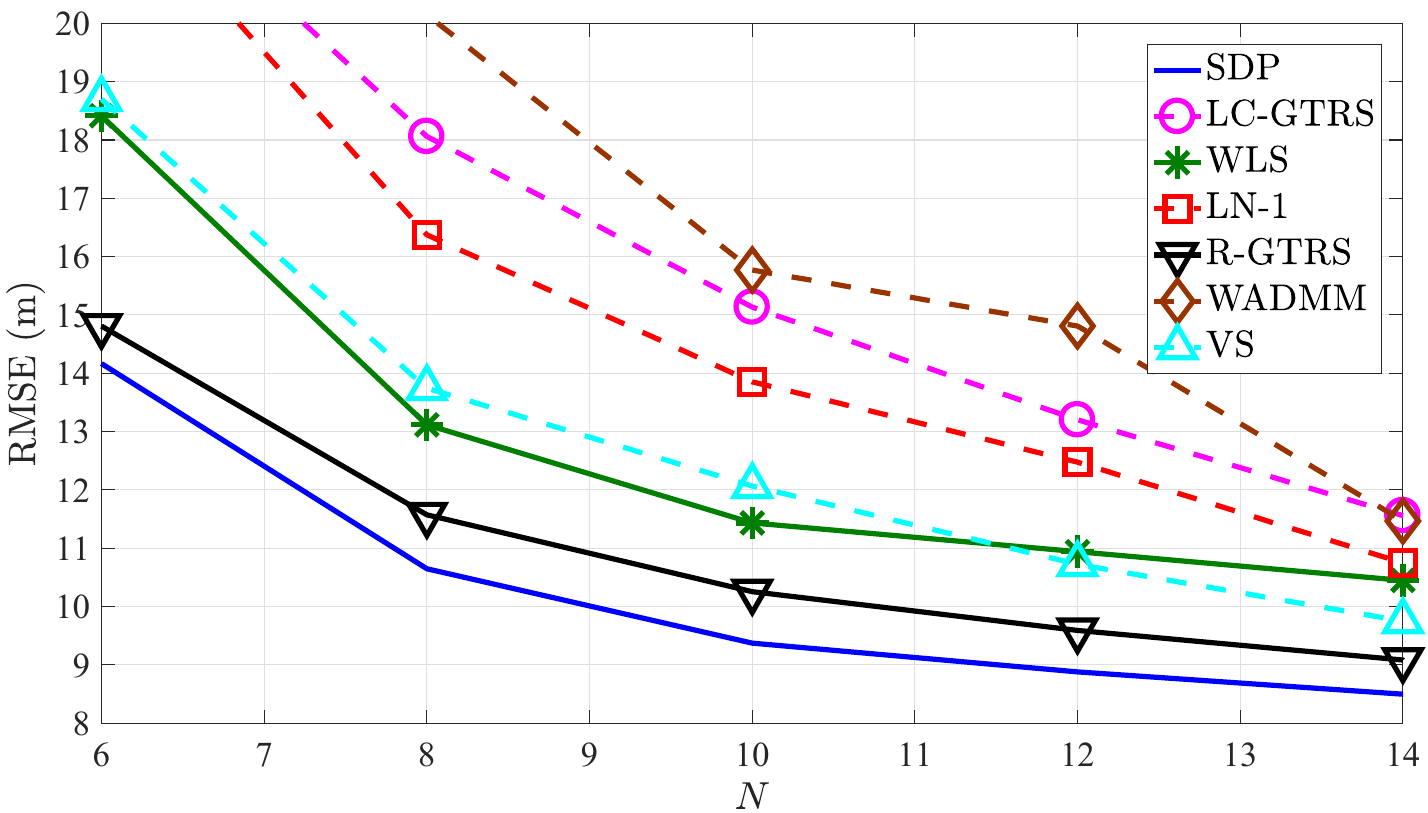}
\caption{RMSE vs. $N$, $\Delta = 20$ and $\sigma = 15$}
\label{fig:RMSE_vs_N}
\end{subfigure}
\begin{subfigure}{.325\textwidth}
\hspace*{-0mm}\includegraphics[width=\textwidth]{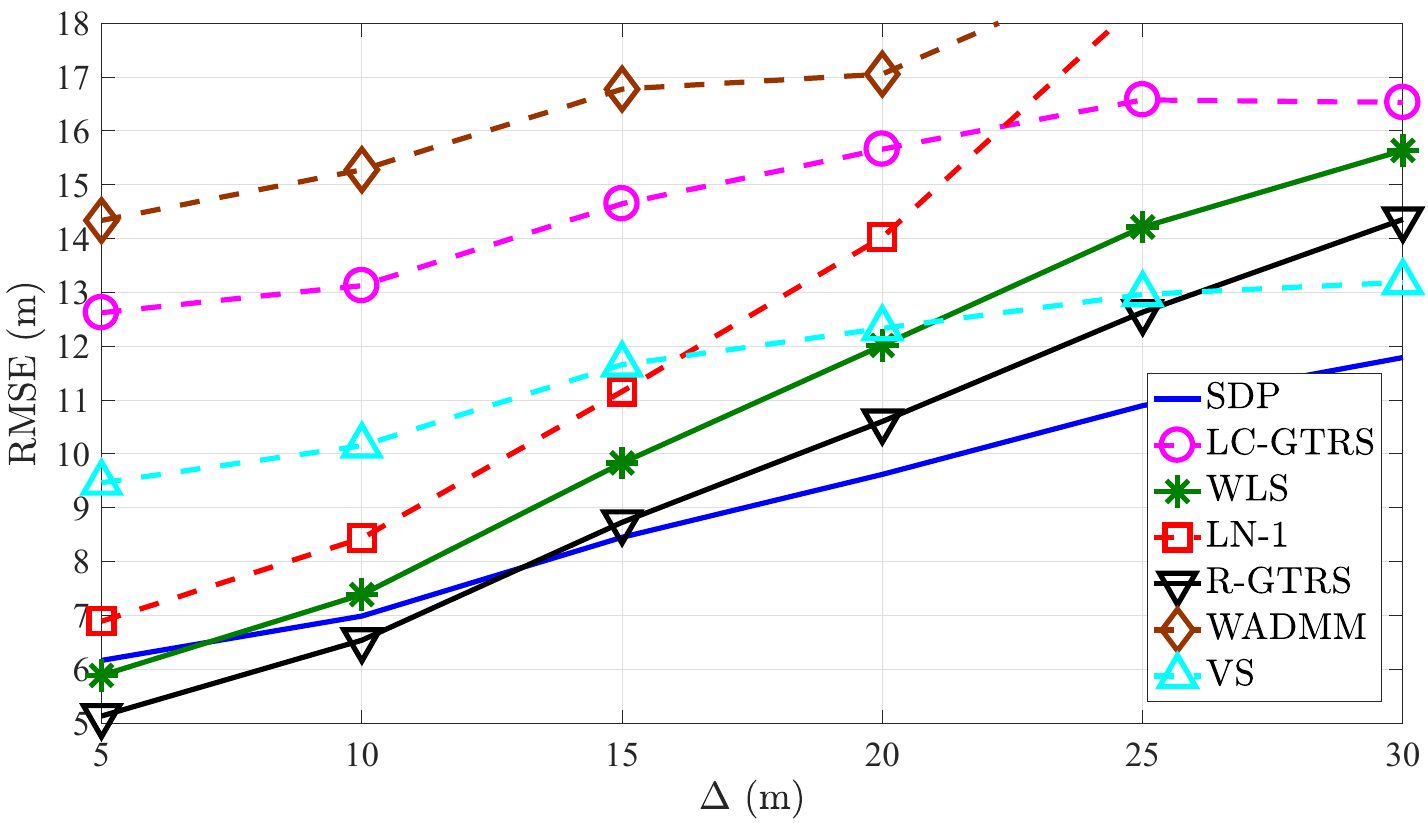}
\caption{RMSE vs. $\Delta$, $N = 10$ and $\sigma = 15$}
\label{fig:RMSE_vs_delta}
\end{subfigure}
\begin{subfigure}{.325\textwidth}
\hspace*{-0mm}\includegraphics[width=\textwidth]{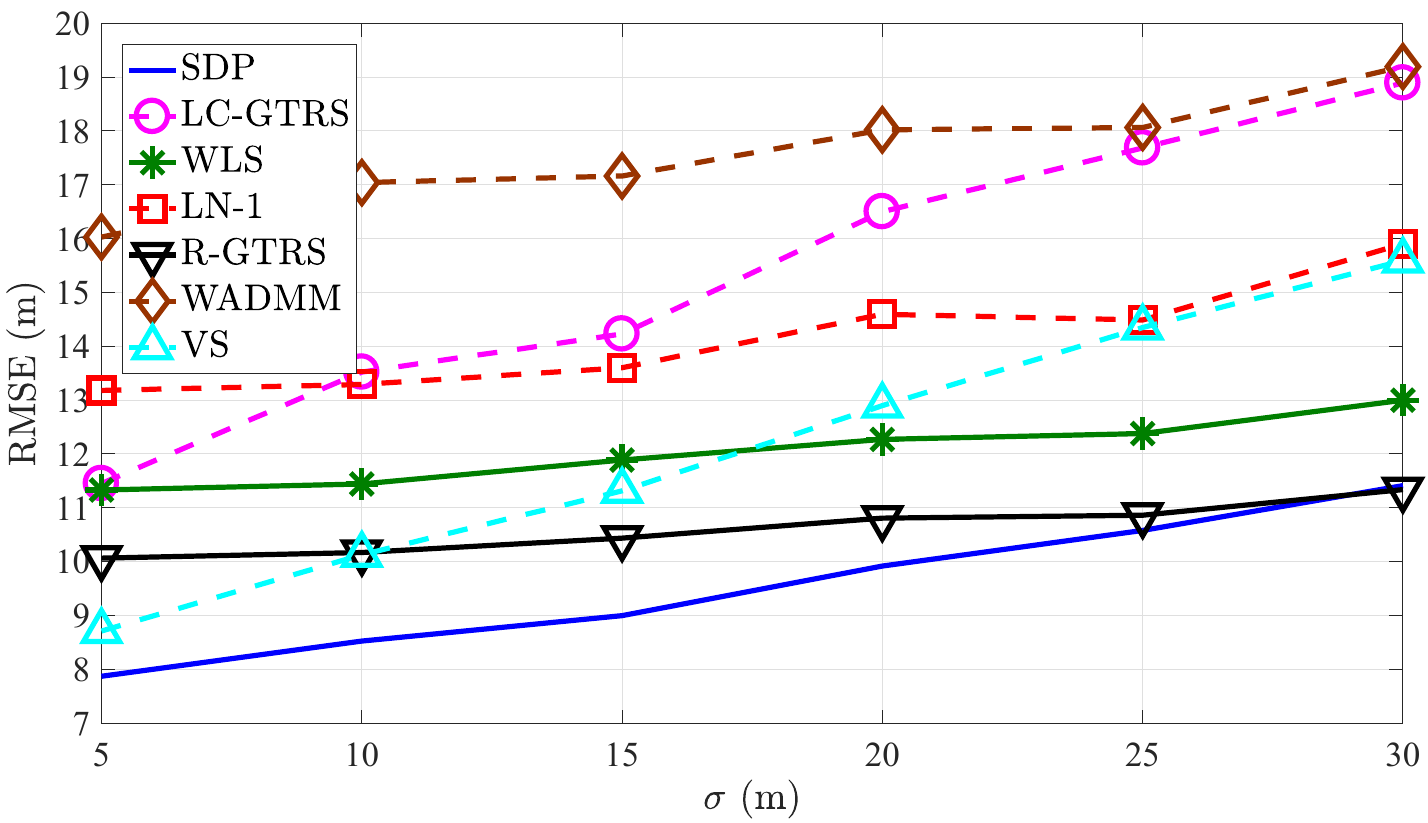}
\caption{RMSE vs. $\sigma$, $N = 10$ and $\Delta = 20$}
\label{fig:RMSE_vs_sigma}
\end{subfigure}
\begin{subfigure}{.325\textwidth}
\hspace*{-0mm}\includegraphics[width=\textwidth]{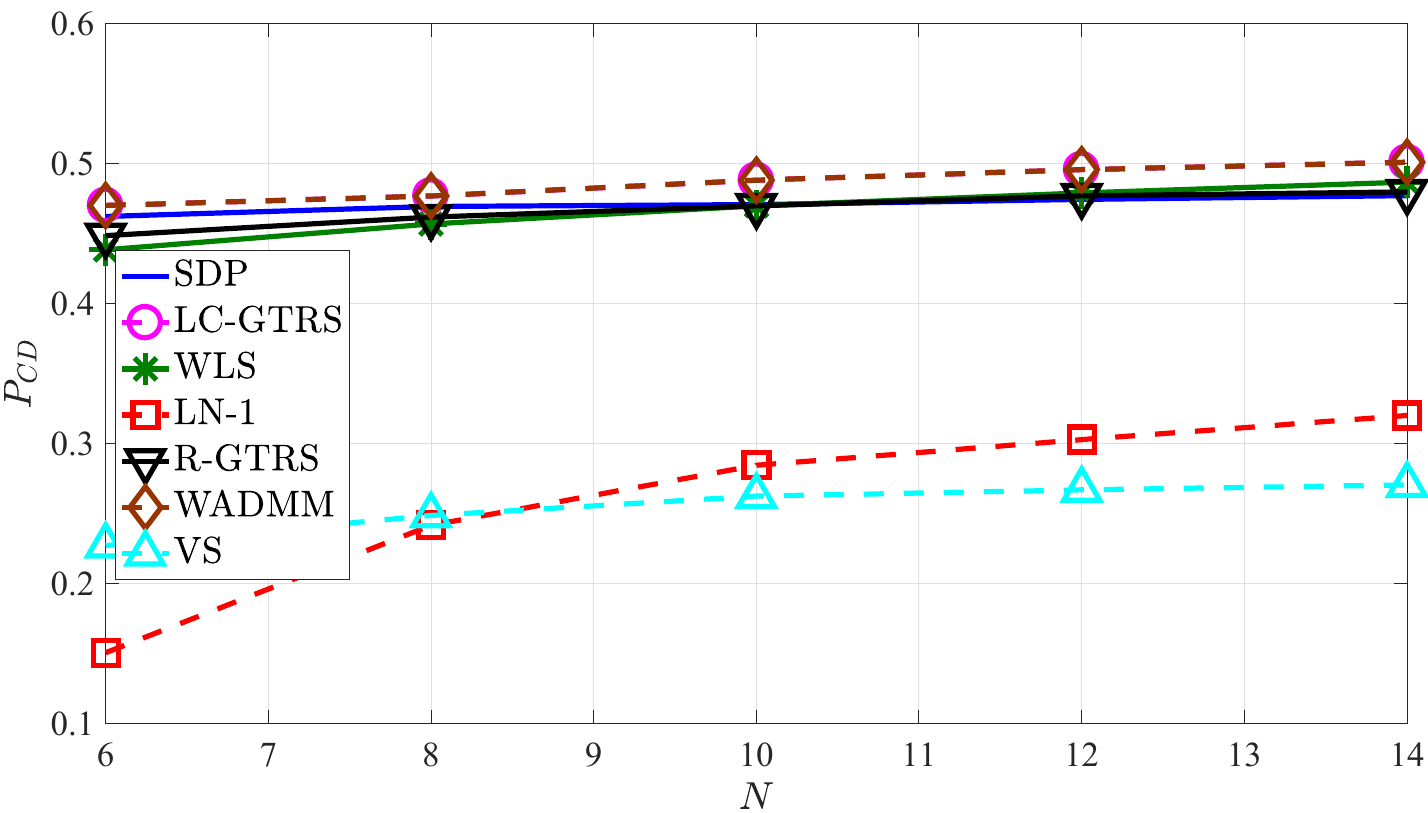}
\caption{$P_{CD}$ vs. $N$, $\Delta = 20$ and $\sigma = 15$}
\label{fig:P_CD_vs_N}
\end{subfigure}
\begin{subfigure}{.325\textwidth}
\hspace*{-0mm}\includegraphics[width=\textwidth]{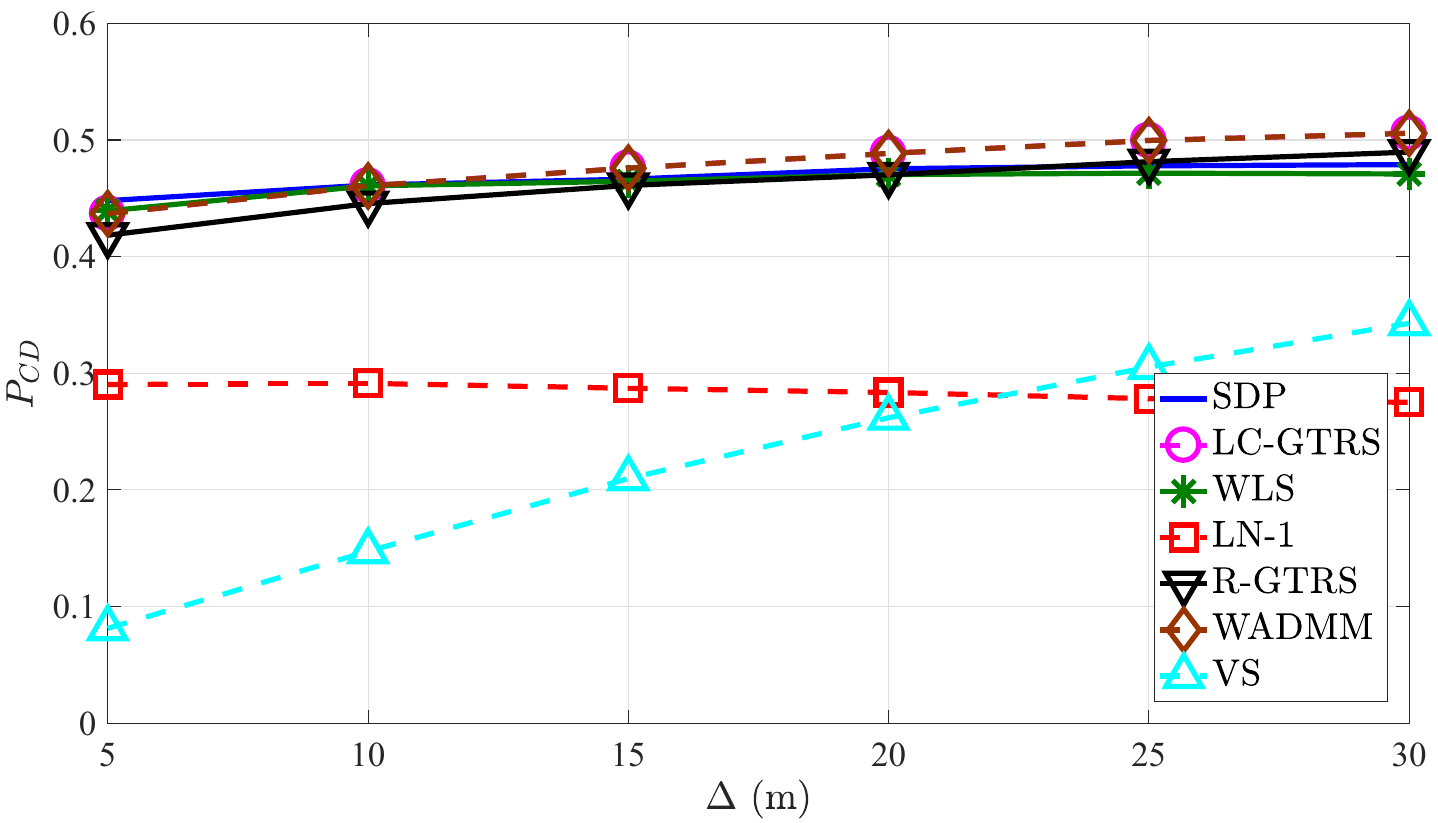}
\caption{$P_{CD}$ vs. $\Delta$, $N = 10$ and $\sigma = 15$}
\label{fig:P_CD_vs_delta}
\end{subfigure}
\begin{subfigure}{.325\textwidth}
\hspace*{-0mm}\includegraphics[width=\textwidth]{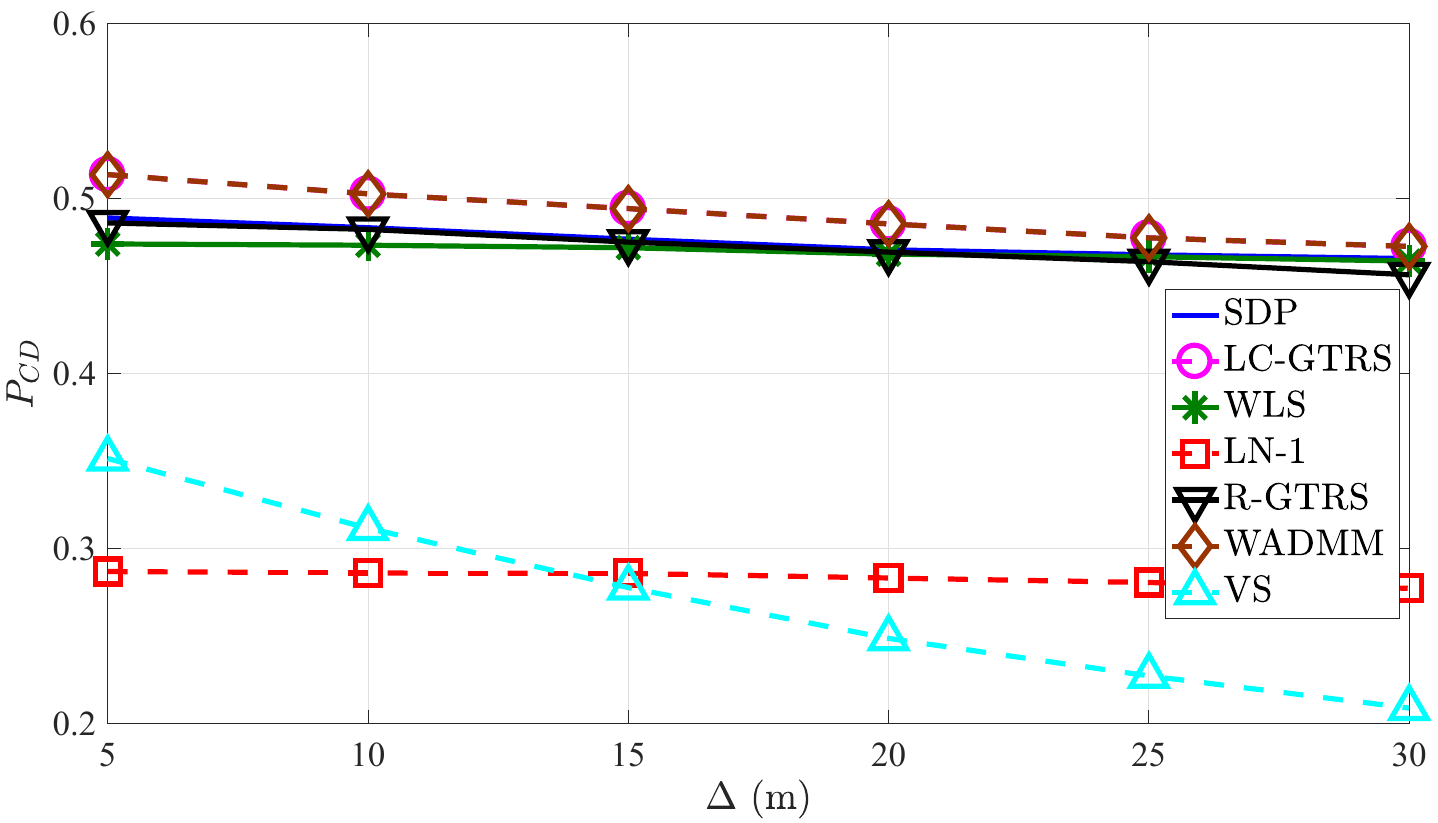}
\caption{$P_{CD}$ vs. $\sigma$, $N = 10$ and $\Delta = 20$}
\label{fig:P_CD_vs_sigma}
\end{subfigure}
\end{center}
\vspace*{0mm}
\caption{Localization and detection performance comparisons in different scenario.}
\label{fig:results}
\end{figure*}

Fig.~\ref{fig:results} validates the performance of the considered algorithms in different scenarios. In terms of localization accuracy, Fig.~\ref{fig:results} shows that the proposed SDP algorithm outperforms the existing ones in general, being R-GTRS the only true competitor in some cases. However, note that R-GTRS in~\cite{Tomic_TVT1:2024} requires additional knowledge about model parameters, namely on the magnitude of the attack intensity, $|\delta_i|$ in~\eqref{eq:model}; thus, the true value of $|\delta_i|$ is given to R-GTRS in all presented simulations. Still, the proposed SDP gains almost $20\%$ in terms of accuracy over R-GTRS when $\Delta = 30$, $\sigma = 15$ and $N = 10$. In terms of detection rates, the figure corroborates the effectiveness of the proposed approach, classifying it among the most superior ones in all considered scenarios. This goes to show that competitive attacker detection can be accomplished even without resorting to sophisticated detection procedures.

\section{Conclusions}
\label{sec:conclusions}

This work introduced a novel approach to reinforce localization credibility in wireless networks in the presence of corrupted/malfunctioning nodes. It proposed neutralizing malicious attacks by considering a surrogate measurement model with dilated noise variance to eliminate malicious attacks by disguising corrupted radio links into ones with large noise variances. In this way, one is able to transform the non-convex MLE into an SDP via a simple CCP procedure. Overall, the proposed method is straightforward to implement and represents a good trade off between localization accuracy and computational complexity, carrying higher computational burden than the existing solutions, but compensating for that by improved localization accuracy. The proposed solution is aimed towards mission-critical applications that require accurate and trustworthy location information and have abundant computational resources available. It also serves to show that the current lower bound on the achievable localization performance can still be brought down.



\bibliographystyle{IEEEtran}
\bibliography{References}

\end{document}